# Game mechanics for cyber-harm awareness in the metaverse

Sophie MCKENZIE[a,1] , Jeb WEBB and Robin DOSS
[a] *Deakin Cyber Research and Innovation*
ORCiD ID: Sophie McKenzie https://orcid.org/0000-0001-5803-640X

**Abstract.** Educating children and young people to be safe online is essential, especially as the metaverse—a next-generation internet blending immersive technologies—promises to reshape their interactions and amplify their experiences. While virtual reality offers fully immersive, highly interactive, and multi-sensory engagement, it also heightens cyber harm risks for young or vulnerable users. To address this, the CyberNinjas VR experience was developed to educate children aged 8 to 16 on safe metaverse behaviours, providing clear referral steps for harmful interactions. Understanding user engagement in metaverse gaming will aid the design of future VR environments which prioritize safety and inclusivity. This project analyses CyberNinjas to understand how game mechanics can foster cyber-safe behaviours.

## 1. Introduction

The metaverse refers to a shared, virtual space that combines physical reality with augmented and virtual realities. It is characterized by persistent digital environments where users can interact through avatars, socialize, and engage in various activities. This transformative space has gained popularity for its potential to revolutionize education, collaboration, and entertainment [1]. Games within the metaverse offer unique opportunities to engage users through immersive and interactive experiences. Effective game design integrates educational objectives with user-friendly interfaces to maximize accessibility and engagement [2]. However, the immersive and interconnected nature of the metaverse also introduces significant challenges, particularly in ensuring user safety and addressing ethical concerns [3]. Cyber harm can occur in this metaverse, and includes technical threats (cyberthreats), content threats (harmful material), and behavioral threats (harmful personal behaviour or social interactions). Generally, the harm associated with the term "cyber harm" is a negative psychological or physical outcome [4].

To help young people avoid experiencing harm in the metaverse, this paper will explore how a cyber harm intervention called CyberNinjas helped young people build cyber awareness. CyberNinjas leverages narrative-driven gameplay and age-specific clue mechanisms to address educational goals of increasing cyber security awareness. Specifically, this paper address; What metaverse specific game mechanics can help inform cyber harm awareness? To address this research question, this paper will describe how the user experience for the game CyberNinjas presented in Virtual Reality (VR) was mapped to cyber harms and supports the user through gameplay.

---

[1] Corresponding Author: Sophie McKenzie, Sophie.mckenzie@deakin.edu.au

**2. Literature: Cyber Harm and the Metaverse**

Cyber harm refers to the negative consequences, either intentional or unintentional—that arise from online activities and interactions within digital environments [5]. It encompasses a wide range of adverse effects that may affect individuals, groups, or organizations, and can manifest in various dimensions, including psychological, physical, economic, social, and reputational harm. These harms often occur due to the interconnected nature of digital spaces, where actions and interactions can form complex chains of cause and effect, making these sources of harm challenging to isolate and address. According to Agrafiotis et al. [5], organizational cyber harm can be categorized into five key types: Physical/Digital; Economic; Psychological; Reputational; Social/Societal. At its core, cyber harm involves both the causes and outcomes of online behaviours and incidents. Causes may include abusive actions (e.g., cyberbullying, harassment, exposure to harmful content), negligent behaviours (e.g., failure to protect privacy, sharing misinformation), and technical threats (e.g., hacking, theft of digital assets) [6]. Outcomes may range from mental health issues, such as anxiety, depression, and distress; to physical consequences like ergonomic injuries; to broader societal impacts like discrimination and reputational damage [6]. To progress our understanding of cyber harm in the metaverse context, we developed a user-centric taxonomy of cyber harm in the metaverse [6], with key characteristics of this taxonomy described below.

*2.1. Key Characteristics of Cyber Harm for Online Experiences*

The taxonomy as described in [4] identifies 17 core harmful behaviours that can take place during any type of online user experience: Assaulting; Criticizing; Defaming; Defrauding; Disclosing; Discriminating; Enabling; Exposing; Harassing; Humiliating; Ignoring; Injuring; Misguiding; Misinforming; Overwhelming; Stealing; and Vandalizing. This taxonomy provides a useful framework for considering online behaviours that may result in harm [4]. However, it is important to note that cyber harm is highly subjective, with individual experiences influenced by personal resilience, context, and psychological makeup. The core behaviours can be intentional (e.g., cyberstalking, doxxing) or unintentional (e.g., accidental disclosure of private information, exposure to harmful content due to inadequate controls). There is also an amplification of effects which can occur through the constant and pervasive nature of modern digital connectivity, where information can spread globally in an instant, making containment or mitigation challenging.

*2.2. Games for Cyber Security Awareness*

Game mechanics in VR can strengthen cyber harm awareness by creating interactive, engaging, and realistic learning environments [7]. These mechanics not only educate users but also provide experiential training that enhances knowledge retention and application in real-world scenarios. Adaptive learning environments in VR can enhance cyber harm awareness by tailoring the difficulty and complexity of scenarios to the user's proficiency level [8]. AI-driven dynamic content adaptation is another powerful tool in VR game design for cybersecurity education [8], as they can adjust the difficulty of cyber threat scenarios to provide personalized learning experiences [9]. Community-driven awareness initiatives within the metaverse can encourage users to share knowledge, report suspicious activities, and collaboratively develop best practices for safer online interactions. These experiences enhance users' ability to identify malicious behaviours and adopt safer practices when navigating digital environments. Incorporating collaborative multiplayer scenarios can also be beneficial [7] to show how users can work together to identify and mitigate threats.

## 3. User Experience of Cyber Harms in CyberNinjas

To explore cyber harms in the metaverse, this study designed and developed a virtual reality (VR) game called CyberNinjas. The game follows the story of Yuki, a young girl who is wrongly accused of harmful behaviour in the virtual world. Players take on the role of investigators tasked with uncovering the truth to reveal that Yuki's account was compromised through social engineering by another child, Lex. As part of their investigation, players discover clues by selecting and inspecting items such as a USB drive (to reveal chat logs), a photo of a lion (to reveal streaming footage), and a chess piece (to reveal clues about avatar changes). The game is tailored to different age groups, with primary-aged mode (8 to 12 years) allowing up to 4 clues and secondary-aged mode (13 to 16 years) offering up to 6 clues for deeper exploration.

When designing and developing CyberNinjas we acknowledge that young people develop cognitively, socially, and emotionally at different rates. We analysed the popularity of games and characters among children to see what themes resonate, such as adventure, fantasy, or educational content, and to better understand what features they in different age groups they enjoy in games. We found that children aged 8-11 need less on-screen text and respond better to bright visuals, while those aged 12-16 can manage more complex interactions and narratives. We prepared CyberNinjas to be harm adjacent, meaning that the player does not enact or receive cyber harm directly. Instead, we used the affordances of VR that promote exploration with an investigation genre of gameplay, to enable players to act as an observer/investigator. Each player explores the environment to understand Lex's actions. Table 1 maps the harms shown in CyberNinjas against the 17 core harm categories as described in [6]. Primarily the harms fall in CyberNinjas under the categories of Abuse, Negligence, and Theft, enabling harms to be covered across numerous categories (as annotated with -1 or -2).

**Table 1.** 17 harm categories of Webb at al., mapped to the CyberNinjas scenario

| Category | Type of Harm | Scenario Element |
|---|---|---|
| **Abuse** | Assaulting | Not applicable |
|  | Criticising-1 | Not applicable |
|  | Disclosing-1 | Lex revealed embarrassing pictures of Yuki |
|  | Exposing-1 | Lex showed others scary pictures that they did not want to see |
|  | Harassing | Lex persistently created problems for Yuki |
|  | Humiliating-1 | Lex intentionally made Yuki feel ashamed by showing embarrassing pictures |
|  | Overwhelming-1 | Not applicable |
|  | Vandalizing | Not applicable |
| **Discrimination** | Discriminating | Not explicit but possible |
| **Misrepresentation** | Defaming | Lex's actions damaged Yuki's reputation by making others believe she was mean |
|  | Misguiding | Lex tricked Yuki into giving him information and misled others by pretending to be Yuki |
|  | Misinforming | Lex used Yuki's account/ avatar to suggest he was her |
| **Negligence** | Enabling | Not applicable |
|  | Criticising-2 | Not applicable |
|  | Disclosing-2 | Lex shared embarrassing pictures of Yuki without her permission |
|  | Exposing-2 | Lex showed scary pictures to other kids without considering their capacity to be disturbed |
|  | Humiliating-2 | Not applicable |
|  | Ignoring | Not applicable |
|  | Injuring | Not applicable |
|  | Overwhelming-2 | Not applicable |
| **Theft** | Defrauding | Not applicable |
|  | Stealing | Lex took Yuki's account credentials without permission |

## 4. Methods

To understand what game mechanics can help inform cyber harm awareness in the metaverse this study engaged in a quantitative approach for data collection and analysis. Our research design used a non-experimental approach to gather research outcomes, gathering in-game data to posit what game mechanics can engage in building cyber harm awareness [10]. A quantitative approach was chosen as it enabled in-game and headset sensor data to inform an understanding of gameplay as drawn during the experience, rather than as a post self-reported activity. Key metrics, including session counts, play time, clue discoveries, and object interactions, were extracted and categorized by age groups into primary (8-11) and secondary (12-16) modes. We analysed our data using descriptive statistics (frequency) to report on clue interaction. The headset and game data did not focus on the nuances of different individual learner approaches to gameplay; rather, aggregated data drawn across sessions was used to gain an overall understanding of player preferences for in-game objects and gameplay options, enabling us to identify areas for potential future improvement [11], [12]. We used non-probabilistic sampling [13], recruiting study participants through schools in Victoria and NSW in Australia who were interested in receiving a CyberNinjas incursion. Each incursion ran for 40 to 50 mins and hosted on average 15 students per session. In total, we engaged with 1057 children aged 8–16 across 12 schools (8 primary schools and 4 secondary schools) introducing them to CyberNinjas. Of the 1057 students, 73% came from primary schools

(aged 8 to 11), with 27% from secondary school (aged 12-16). Ethics approval code: DHREC-2023-371, was obtained from the authors institution.

## 5. Results

The average time it took for each student to complete a game of CyberNinjas across both the primary and secondary school group was 15 minutes, with almost 100% completion of the game by each student. On average during an in-school incursion, the CyberNinjas game was played 23 times in full. We found that students who completed the primary-aged experience would then attempt the more complex secondary-aged version. The average number of game replays by each student was 2, however it was not uncommon for students to play the game more than 2 times in a 50-minute incursion session. Table 2 includes a count of the numbers of times a student interacted with a clue or object during a game of CyberNinjas. A rough percentage, rounded to the nearest whole number, has also been included in Table 2. Clue 5 and 6 only appeared in the secondary school version of CyberNinjas, therefore with removing clue 5 and 6 the total number of clues in the secondary experience is 3142.

**Table 2.** Interaction with clues or objects in CyberNinjas

|  | **Primary Mode (Aged 8-11)** | **Secondary Mode (Aged 12-16)** |
| --- | --- | --- |
| **Total number of interactions** | *4288 interactions total* | *4260 (3142 with clue 5 and 6 removed)* |
| **Interacted with USB Drive** | 702 (17%) | 562 (13%) |
| **Interacted with Fish Food** | 682 (16%) | 535 (13%) |
| **Interacted with Photo of Lion** | 577 (14%) | 440 (10%) |
| **Interacted with Chess Piece** | 685 (16%) | 524 (13%) |
| **Interacted with Toy Plane** |  | 578 (14%) |
| **Interacted with Prize** |  | 540 (13%) |
| **Interacted with Coffee Machine** | 109 (2.5%) | 67 (2%) |
| **Interacted with Basketball** | 992 (21.5%) | 670 (16%) |
| **Interacted with Paper Shredder** | 32 (1%) | 24 (1%) |
| **Interacted with Donut** | 509 (12%) | 320 (8%) |

Table 3 provides an average of the number of object interactions that occurred when a student completed a game of CyberNinjas. There were some differences in the number of object interaction, with the photo of the lion (as an essential clue to complete the experience) being interacted with less than expected.

**Table 3.** Average number of clue or object interactions in CyberNinjas

| USB Drive | Fish Food | Photo of Lion | Chess Piece | Toy Plane | Prize |
| --- | --- | --- | --- | --- | --- |
| 43.88 | 42.33 | 36.36 | 42.03 | 23.33 | 21.67 |

*5.1. Metaverse games mechanics for cyber harm awareness*

In CyberNinjas, using the USB reveals chat logs showing where Yuki shared personal details, like her pet's name with Lex, disclosing information. Lex was then able to manipulate Yuki's identity and gain access to her account. This clue is not a game mechanic specific to the metaverse, however thematically it progressed the CyberNinjas investigative scenario. The USB clue received higher on average object interactions (table 3), highlighting players may not have known how to complete the USB clue. Other clues as a part of the game mechanics for CyberNinjas included affordances that relate to the metaverse context. Feeding the fish reveals that Lex bragged about impersonating Yuki and manipulating her avatar, misguiding on online profile. The immersive context of VR amplifies impersonation, with impersonation of voice, appearance, mannerisms, etc highlighted in the metaverse. A photo confirms in CyberNinjas that Yuki was playing basketball at 2pm, proving she was not playing football, showing Lex defaming and harassing Yuki. While defamation and harassment have always been possible online, an immersive environment makes this possible in ways which are potentially more intense and invasive. The photo of the lion provided important clue information yet, on average, was interacted with less. This indicates that players may have found this clue easier to solve. The chess piece clue reveals that Yuki's avatar had been changed, suggesting unauthorized access. Like the other clues, while impersonation has always been possible online, an immersive environment makes this possible in dynamic, real-time ways that are potentially more engaging, convincing or persuasive when in the metaverse. Throwing the toy plane reveals that Lex shared embarrassing pictures of Yuki with other players, exposing and humiliating her. Finally, the trophy clue reveals that Yuki shared her full birthday details with a stranger to win a prize, making her vulnerable to identity theft and further exploitation. While many of the harms explored in CyberNinjas could be learnt about in other ways [7], an immersive environment can amplify how appealing or engaging a person is, how safe an environment feels, how urgent a situation is, or many other factors which contribute to the effectiveness of a social engineering attempt [9].

## 6. Conclusion

Through activities like investigating clues, interacting with objects, and reflecting on harm scenarios, CyberNinjas provides a unique platform to explore the complexities of cybersecurity in the metaverse. The results show how game mechanics can be applied to build awareness of cyber harm in the metaverse, particularly when using a game-based environment to support young people [9]. The study leveraged our user-centric harms taxonomy [6] to examine the game mechanics, and VR- specific engagement tools in fostering cyber awareness resilience against online risks. Our mapping indicates challenges particular to the metaverse, showing how immersive aspects may impact how cyber harms are portrayed. The insights gained from this research offer a foundation for designing future VR experiences that balance engagement with protection, helping to create metaverse environments that support leaners to safely explore important topics.